\documentclass[preprint,12pt, a4paper]{elsarticle}

\usepackage{amssymb}
\usepackage{comment}
\usepackage{hyperref}
\usepackage{booktabs}
\usepackage{fontawesome5}
\usepackage{enumitem}
\usepackage{float}
\usepackage[table]{xcolor}
\usepackage{colortbl}
\usepackage{tikz}
\usepackage{pgfplots}
\usepackage{standalone}
\usepackage{iftex}
\ifLuaTeX\usepackage{emoji}\fi
\usepgfplotslibrary{groupplots}
\usetikzlibrary{arrows.meta,positioning,fit}
\pgfplotsset{compat=1.18}

\definecolor{lightgray}{gray}{0.85}
\journal{SoftwareX}

\begin{document}
\renewcommand{\labelenumii}{\arabic{enumi}.\arabic{enumii}}

\begin{frontmatter} 
 


\title{QGas: Interactive Gas Infrastructure Toolkit}

\author[1,2]{M. Quantschnig}
\author[1,2]{Y. Werner}
\author[1,2]{S. Wogrin}
\author[1,2]{T. Klatzer\corref{correspondingauthor}}
\cortext[correspondingauthor]{Corresponding author:}
\ead[url]{thomas.klatzer@tugraz.at}

 \affiliation[1]{organization={Institute of Electricity Economics and Energy Innovation (IEE), Graz University of Technology},
            addressline={Inffeldgasse 18}, 
            city={Graz},
            postcode={8010},
            country={Austria}}

 \affiliation[2]{organization={Research Center ENERGETIC, Graz University of Technology},
             addressline={Rechbauerstraße 12},
             city={Graz},
             postcode={8010},
             country={Austria}}

\begin{abstract}
Gas infrastructure datasets are essential inputs for energy system planning to support strategic decision-making toward decarbonization. However, relevant data are typically scattered across heterogeneous sources, including geospatial datasets, image-based infrastructure plans, and tabular data, making it complex, time-consuming, and error-prone to create topology-consistent network representations with existing tools.
This paper presents QGas, an interactive toolkit for visualizing, creating, and collaboratively extending georeferenced gas infrastructure datasets. QGas integrates GIS-based geometry editing with topology-preserving graph operations in a unified web-based environment, enabling users to digitize infrastructure plans, edit network elements, manage attributes, and perform topology-consistent modifications while maintaining a georeferenced representation of the system.
The toolkit is implemented using a modular architecture based on Python, JavaScript, and the Leaflet mapping library. An illustrative example demonstrates its application in extending a natural gas dataset to include hydrogen and CO\textsubscript{2} infrastructure, highlighting QGas’s capability to support the preparation of consistent multi-carrier gas infrastructure datasets for energy system planning.
\end{abstract}


\end{frontmatter}

\section*{Metadata}
\label{}

\begin{table}[H]
\begin{tabular}{|l|p{6.5cm}|p{6.5cm}|}
\hline
\textbf{Nr.} & \textbf{Code metadata description} & \textbf{Metadata} \\
\hline
C1 & Current code version & v1.0 \\
\hline
C2 & Permanent link to code/repository used for this code version & \url{https://github.com/IEE-TUGraz/QGas}/releases/tag/v1.0  \\
\hline
C3  & Permanent link to Reproducible Capsule & - \\
\hline
C4 & Legal Code License   & MIT License \\
\hline
C5 & Code versioning system used & git \\
\hline
C6 & Software code languages, tools, and services used & JavaScript, Python; HTML, CSS; Tkinter (GUI), Pillow; MkDocs (documentation) \\
\hline
C7 & Compilation requirements, operating environments \& dependencies & Python 3.11; Conda environment setup via environment.yml; Python dependencies include Pillow; Tkinter required (bundled with Python); a modern web browser required to run the JavaScript-based GUI (GUI.html)\\
\hline
C8 & If available Link to developer documentation/manual & \url{https://qgas.readthedocs.io/} \\
\hline
C9 & Support email for questions & thomas.klatzer@tugraz.at \\
\hline
\end{tabular}
\caption{Code metadata}
\label{codeMetadata} 
\end{table}

\section{Motivation and Significance}
\label{sec:Intro}
The transition towards decarbonized energy systems places growing emphasis on integrating the power and gas sectors~\cite{EUEnergySystemIntegration2020} to accommodate large volumes of alternative gases, such as renewable hydrogen~\cite{EURFNBO2023}, or CO\textsubscript{2}. This integration is expected to transform today’s gas infrastructure into multi-carrier systems that evolve continuously and exhibit changing topological characteristics over the coming decades. Planning and operating such complex networks is commonly supported by open-source energy system optimization models, such as PyPSA~\cite{PyPSA2018}, ZEN-garden~\cite{Mannhardt2025}, GENeSYS-MOD~\cite{Loeffler2017}, Tulipa~\cite{Tulipa2023}, or LEGO~\cite{WOGRIN2022101141}, which rely on coherent, accurate, and up-to-date infrastructure information.

Against this background, preparing comprehensive gas infrastructure datasets is an essential prerequisite for reliable energy system modeling. Yet this task remains challenging, as relevant information is dispersed across various formats and a wide range of sources~\cite{McWilliams2025Energy}. Generally, existing data sources can be grouped into three categories: (i) graph-based datasets that provide topologically consistent network representations, such as SciGRID\_gas~\cite{Pluta2022} and GasLib~\cite{GasLib2017}; (ii) sources without explicit graph representations, such as OpenStreetMap~\cite{OpenStreetMap2025} and Global Energy Monitor~\cite{GEM_Tracker2024}; and (iii) graphical or image-based infrastructure plans, such as the ENTSOG Transparency Map~\cite{ENTSOGMAP2024} or regional network expansion plans, e.g.,~\cite{AGGM2024CO2Study},~\cite{AGGM2024H2Roadmap}, which typically provide only coarse information on technical infrastructure attributes. While these sources collectively offer valuable and complementary information, integrating them into a topologically consistent network representation requires substantial manual effort, including data conversion, georeferencing, topology reconstruction, and modification. These processes are time-consuming, error-prone, and often difficult to reproduce. Moreover, the transformation from single-carrier to multi-carrier gas networks introduces significant structural changes over time that cannot be adequately represented by static datasets alone. As a result, gas datasets must be continuously adapted to reflect evolving system configurations.

Existing tools provide only limited support for such dynamic modifications and data integration tasks. While GIS-based environments such as QGIS~\cite{QGIS2024} and ArcGIS~\cite{EsriArcGIS2024} enable detailed geometric editing and the integration of heterogeneous spatial data, they do not inherently preserve network topology~\cite{ijgi3020662}. Conversely, graph-based tools, e.g.,~\cite{Gephi2024}, \cite{Shannon2003Cytoscape}, \cite{Mueller2020pandapipes}, ensure structural consistency but typically lack integrated geospatial editing capabilities and direct support for heterogeneous data sources. This separation complicates both the integration of diverse data sources and the creation of consistent, up-to-date gas infrastructure datasets for energy system modeling.

Given the limitations of existing tools, ensuring a consistent network topology during interactive dataset modifications presents a key challenge. Even seemingly simple operations, such as splitting a node into multiple subnodes to decouple different carrier systems (e.g., natural gas pipelines and repurposed hydrogen pipelines), require several coordinate changes in the underlying graph structure. In conventional GIS-based workflows, this entails manually splitting connected edges, creating new nodes, reassigning connections, and performing explicit topology validation. As these steps are not inherently linked, inconsistencies such as disconnected components or invalid references can easily occur.

We address these limitations by introducing the QGas toolkit, which provides an environment for combining heterogeneous data sources and performing topology-aware editing operations. Specifically, QGas introduces three key novelties:
\begin{itemize}
    \item It enables the incorporation of heterogeneous data sources such as graph-based datasets, geospatial data without explicit topology, image-based infrastructure plans, and tabular information into a unified, georeferenced network representation.
    \item It combines user-guided interaction with fully automated graph-preserving transformations to ensure topological consistency during dataset modification. For example, when splitting a node, the user specifies the number of resulting subnodes and assigns connected pipelines via graphical interaction, while the system automatically performs all necessary reference updates and preserves the underlying network structure.
    \item It facilitates the visualization, exploration, creation, maintenance, and collaborative extension of datasets for continuously evolving multi-carrier gas systems, thereby enhancing their transparency, consistency, and reproducibility for energy system planning.
\end{itemize}

The remainder of this paper is organized as follows: Section~\ref{sec:Software} describes the QGas software and its main functionalities. Section~\ref{sec:Example} presents an illustrative example in which an existing natural gas dataset is extended by planned hydrogen and CO\textsubscript{2} infrastructure, enabling integrated multi-carrier analysis. Section~\ref{sec:Impact} summarizes previous and ongoing use cases of QGas, and Section~\ref{sec:Conclusions} concludes the paper.

\section{Software Description}
\label{sec:Software}
This section describes the software architecture of the QGas toolkit. 

\subsection{Software Architecture}
\label{sec:Achitecture}
\begin{figure}[t]
    \begin{center}
    \includegraphics[width=1\linewidth]{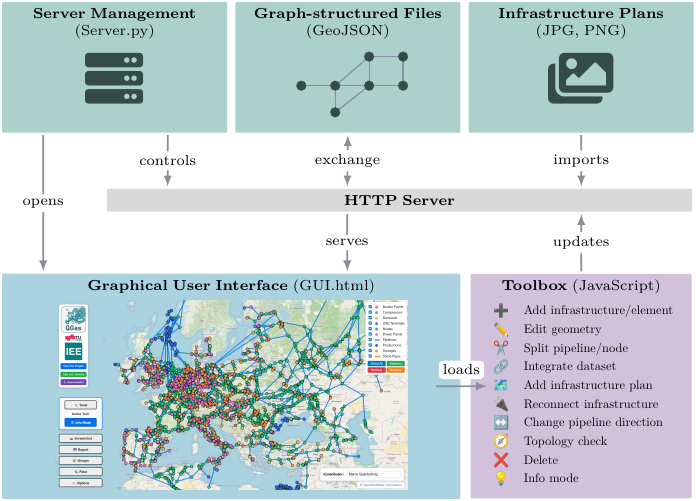}
    \caption{Software architecture and component interaction of QGas.}
    \label{fig:FlowChart}
    \end{center}
\end{figure}

QGas comprises three main components that together form a modular and extensible software architecture. Fig.~\ref{fig:FlowChart} illustrates the QGas software architecture and component interactions. 
A Python-based server control environment (\texttt{Server.py}) handles server management and process control. It hosts a local HTTP server on port 8000 and enables data exchange between graph-structured GeoJSON files and the HTML-based graphical user interface (\texttt{GUI.html}), which provides georeferenced visualization of infrastructure datasets and image-based infrastructure plans. Interactive map rendering is implemented using the Leaflet library~\cite{Leaflet2024}, allowing users to explore and inspect spatial network structures.
A modular JavaScript framework implements the toolbox for user interaction logic and dataset modification. This toolbox manages editing operations, attribute handling, and topology-preserving modifications of the network graph.

The overall software architecture is designed to ensure modularity, maintainability, and extensibility. Individual components can be updated or extended independently, facilitating future development and integration of additional functionalities. 

\subsection{Data Modification and Analysis Toolbox}
\label{sec:JavaScrip}
QGas enables interactive modification and analysis of infrastructure datasets through a modular JavaScript toolbox. This toolbox separates core functionality, shared utilities, user interface components, and specialized tools for editing and analysis. Its modular design supports maintainability, scalability, and long-term development.

\begin{figure}[t]
    \begin{center}
    \includegraphics[width=1\linewidth]{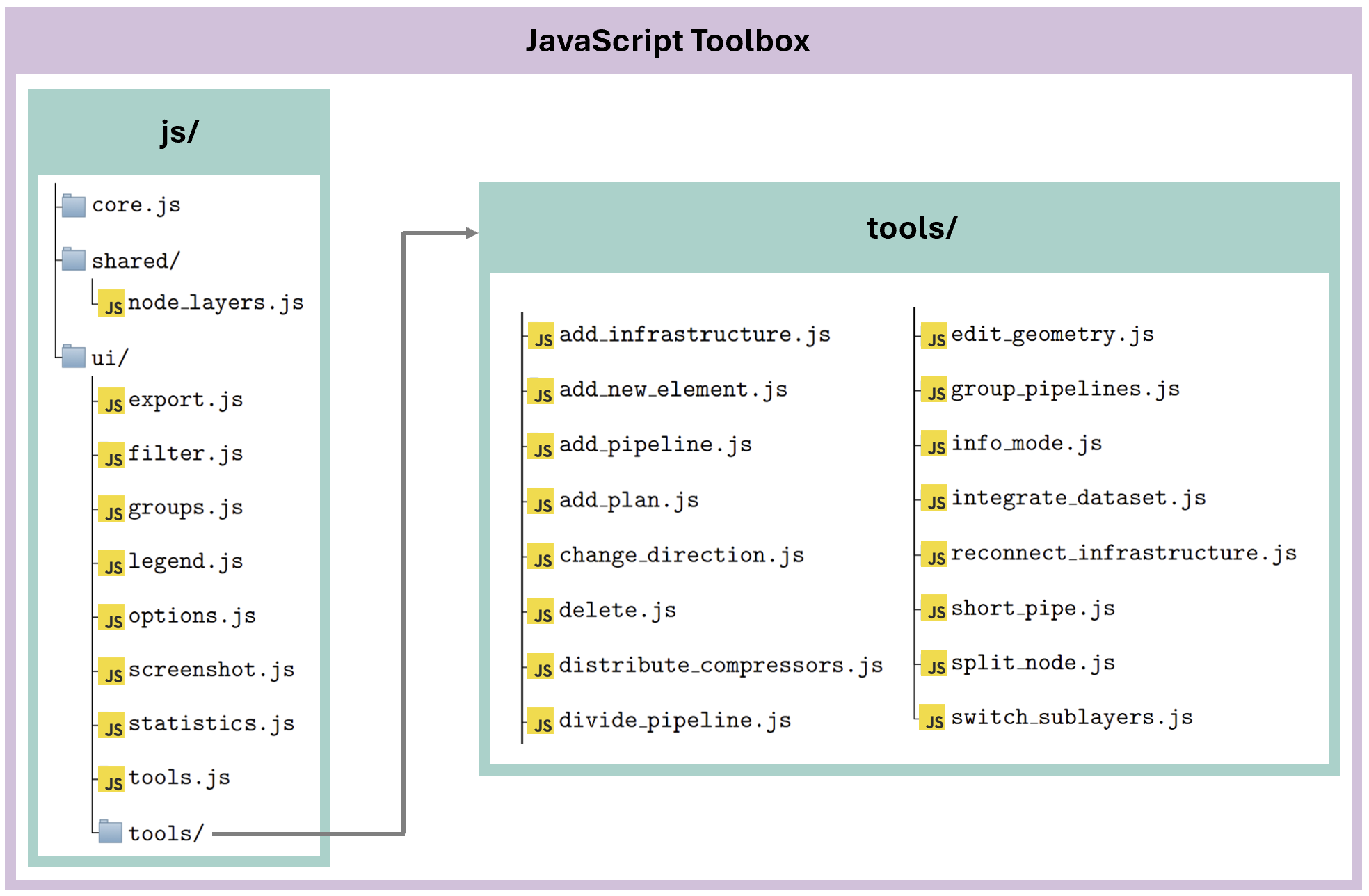}
    \caption{Structure of the JavaScript toolbox framework of QGas.}
    \label{fig:JavaScript}
    \end{center}
\end{figure}

The \texttt{core.js} module provides the fundamental functionality required to load, initialize, and format data layers, and to establish the basic system configuration. Functions used across multiple tools are located in the \texttt{shared} directory. 
The \texttt{ui} directory contains modules for user interface elements and visualization components such as filtering and export logic. A detailed description of the graphical user interface is provided in Subsection~\ref{sec:GUI}.
Dataset operation tools are implemented in the \texttt{tools} directory. Each tool is organized as an independent module following a standardized interface for data access and modification, allowing new functionalities to be added or tools to be extended with minimal impact on the software architecture. The structure of the JavaScript toolbox framework is illustrated in~Fig.~\ref{fig:JavaScript}.

\subsection{Dataset Structure}
\label{sec:DataStructure}
This subsection describes the structure of a dataset project created and managed with QGas.
To facilitate immediate use after installation, QGas provides a reference dataset derived from the SciGRID\_gas IGGIELGNC-1 dataset \cite{Pluta2022,Diettrich2021SciGRIDgas}. This dataset was converted to the QGas project structure and serves as a template for new projects and testing.

The core graph structure of each dataset is reference-based and defined by two mandatory files: \texttt{nodes.geojson} containing all network nodes and \texttt{pipelines.geojson} storing the corresponding network edges. Together, these files form the topological backbone of the dataset. Connections are realized by node references. Each pipeline references a start and an end node within the attribute set.  
Other infrastructure elements, such as compressor stations, storage facilities, and other network components, can be created within QGas using the integrated toolbox functions. Each of these elements contains an attribute-based node reference indicating the node to which it is connected. These elements are exported as separate, component-specific \texttt{layers.geojson} files, enabling the integration of arbitrary infrastructure layers, e.g., hydrogen pipelines, electrolyzers, or biogas facilities, while maintaining compatibility with the core network representation. 
Layer initialization and default visualization settings are defined in the \texttt{conf.xlsx} file. This configuration file specifies legend entries, layer types, and styling parameters, such as marker symbols and color schemes. It is automatically updated when the dataset is exported, ensuring consistency between the configuration and the stored data.
The \texttt{plans} directory stores georeferenced images and infrastructure plans imported into the QGas environment. These images serve as visual references for digitization and network reconstruction.
Licensing and data source information are documented in \texttt{license.tex}. Information entered by the user is displayed in the integrated licensing section of the graphical user interface, ensuring transparent documentation of data provenance and usage rights. The described structure of a QGas dataset project is summarized in Table~\ref{tab:files}.

\begin{table}[!t]
\caption{Structure of a QGas dataset project.}
\label{tab:files}
\centering
\scriptsize
\renewcommand{\arraystretch}{1.15}
\begin{tabular}{lll}
\hline
\textbf{File} & \textbf{Type} & \textbf{Description} \\
\hline
pipelines       & .geojson  & Main layer for topology edges (mandatory)            \\
nodes           & .geojson  & Main layer for topology nodes (mandatory)            \\
layers          & .geojson  & Contains information on other infrastructures        \\
conf            & .xlsx     & Contains default layer types and styling             \\
plans           & directory & Contains georeferenced infrastructure plans          \\
license         & .txt      & Licensing information used within the tool           \\
\hline
\end{tabular}
\end{table}

\subsection{Graphical User Interface}
\label{sec:GUI}
\begin{figure}[t]
    \centering
    \includegraphics[width=1\linewidth]{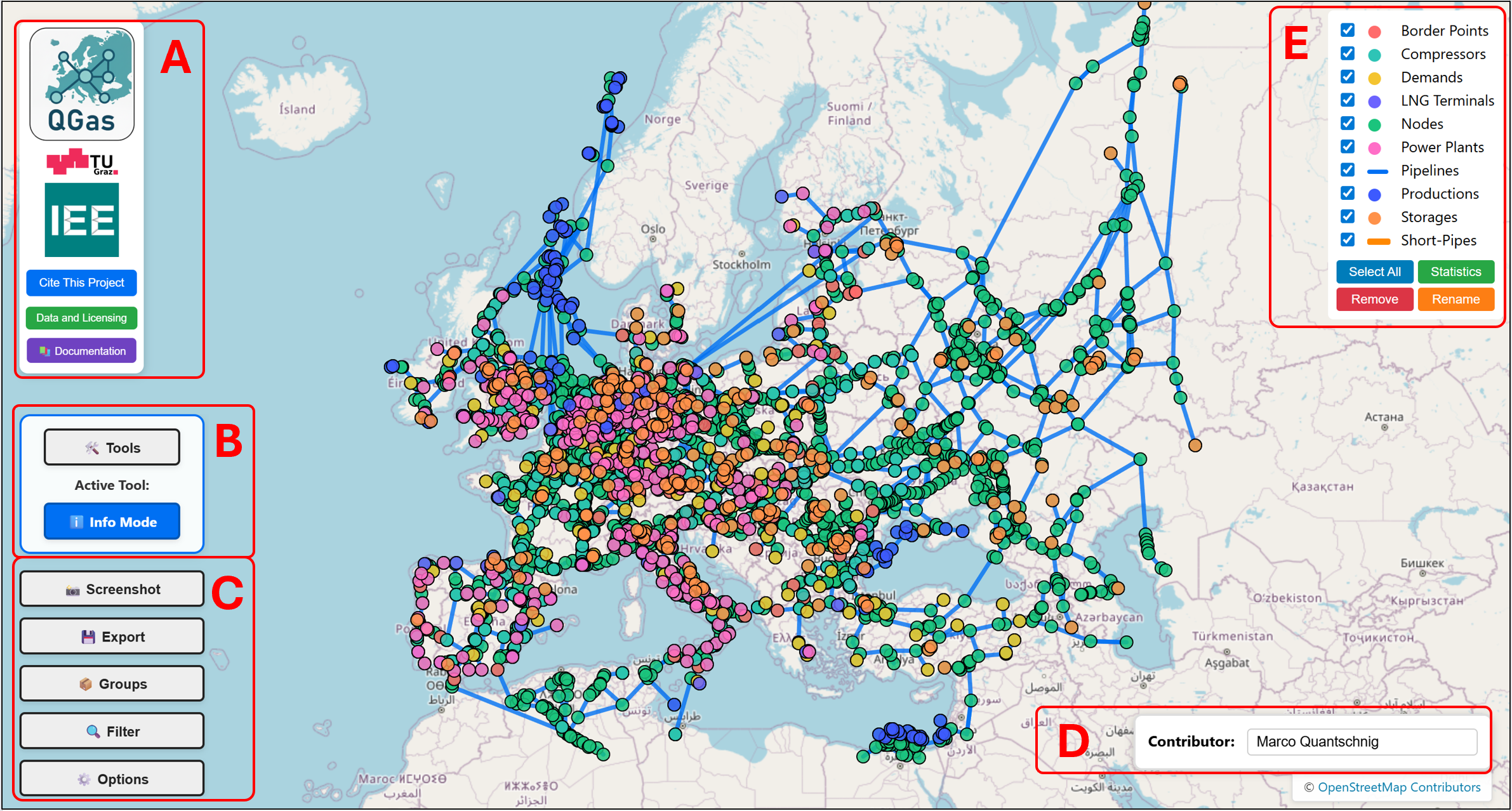}
    \caption{Main screen of QGas.}
    \label{fig:User_Interface}
\end{figure}

This section provides an overview of the graphical user interface of QGas, implemented in \texttt{GUI.html}, which serves as the primary workspace for dataset visualization and interaction. The main screen is shown in Fig.~\ref{fig:User_Interface}. The interface is designed to support intuitive navigation, efficient editing, and transparent documentation of infrastructure datasets. The central map view provides a georeferenced visualization of all active layers and enables direct interaction with network elements. In the following, we summarize the main interface components.

\begin{enumerate}[label=\Alph*.]
  \item \textbf{Data, citation, and documentation}.
    Provides information on data sources, licensing conditions, and citation requirements, as specified in \texttt{license.txt}, to support transparent data use and reproducibility. Links to the project documentation.

  \item \textbf{Tools}.
  Contains all dataset modification and editing tools available in QGas. The currently active tool is highlighted, enabling users to switch efficiently between editing modes.

  \item \textbf{General functionalities}.
  Provides access to global project functions, including dataset export, group management, spatial filtering, visualization styling, and screenshot generation.

  \item \textbf{Contributor tracking}.
  Displays the active user and records dataset modifications. This feature facilitates collaborative work by enabling traceability of edits in multi-user and multi-project environments.

  \item \textbf{Legend and statistics panel}. 
  Lists all active layers and allows users to enable or disable individual components. In addition, it provides live statistics on network characteristics, such as total pipeline length, element counts, and documented data sources.
\end{enumerate}

\subsection{Tools}
\label{sec:Tools}
\begin{figure}[t]
    \centering
    \includegraphics[width=1\linewidth]{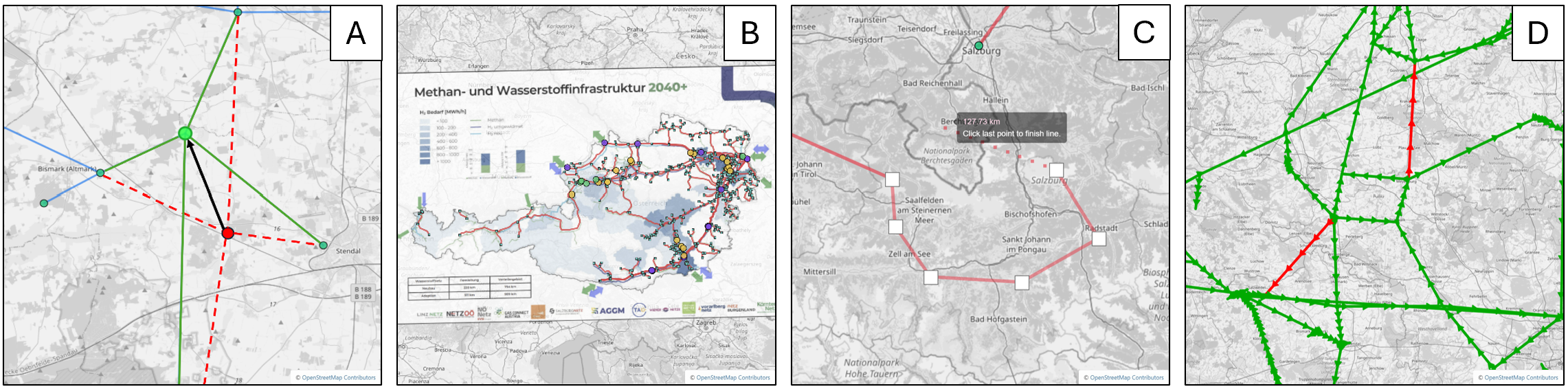}
    \caption{Selection of QGas tools. Repositioning of network nodes (A), georeferencing of image-based infrastructure plans (B), adding new pipelines (C), and setting default pipeline direction (D).}    \label{fig:Tools}
\end{figure}

QGas provides a comprehensive set of interactive tools for editing, extending, and restructuring gas infrastructure datasets. These tools can be grouped into four main categories based on functionality: attribute and geometry editing, network topology operations, layer and infrastructure management, and data integration. A selection of these tools is illustrated in Fig.~\ref{fig:Tools}.

\subsubsection{Attribute and Geometry Editing}
In \textbf{Info Mode}, users can inspect and modify element attributes, such as pipeline name, diameter, or pressure, by selecting network components directly on the map. A flexible attribute management system enables the addition, modification, or removal of attributes across entire layers.
\textbf{Edit Geometry} enables interactive modification of node positions and pipeline routes while preserving the underlying graph topology. \textbf{Delete} removes selected elements for controlled dataset cleanup.

\subsubsection{Network Topology Operations}
Several tools are dedicated to maintaining and restructuring the graph-based network representation. The \textbf{Add Pipeline} tool creates new pipeline segments by defining route points on the map. The geometric length of newly created pipelines is calculated automatically, and predefined attribute fields are assigned.
The \textbf{Divide Pipeline} function subdivides existing pipeline segments and generates intermediate nodes for localized attribute modifications.
The \textbf{Split Node} tool creates multiple subnodes at the location of an existing node, enabling the decoupling of interconnected carrier systems.
The \textbf{Change Direction} tool visualizes and inverts pipeline flow directions.
The \textbf{Short-Pipe} tool supports temporary or structural connections between nodes by introducing lossless connector elements.
The \textbf{Reconnect Infrastructure} tool reassigns infrastructure elements and nodes to new connection points while maintaining graph consistency.
The \textbf{Distribute Compressors} tool distributes compressors across multiple pipeline loops.

\subsubsection{Layer and Infrastructure Management}
The \textbf{Add New Element} tool defines new element types, including line, node, point, and in-line components.
\textbf{Add Infrastructure} places additional infrastructure elements and links them to network nodes.
\textbf{Switch Sublayers} creates attribute-consistent sublayers and allows elements to be reassigned, e.g., for infrastructure repurposing from natural gas to hydrogen.
\textbf{Group Pipelines} logically aggregates pipeline segments and provides summary statistics, such as total group length.

\subsubsection{Data Integration and Referencing}
The \textbf{Add Infrastructure Plans} tool imports image-based infrastructure plans as semi-transparent background layers. Georeferencing is performed by aligning identifiable landmarks with the OpenStreetMap base layer, enabling accurate digitization of planned network elements.
The \textbf{Integrate Dataset} tool imports and harmonizes external geospatial datasets, facilitating attribute transfer between overlapping network sections.

\section{Illustrative Example}
\label{sec:Example}
We demonstrate the capabilities of QGas by modifying a dataset representing the natural gas infrastructure of the Austrian federal state of Carinthia, which is currently subject to several transition plans related to hydrogen deployment and envisioned CO$_2$ infrastructure. These developments provide a suitable example of how existing gas infrastructure datasets can be extended to multi-carrier systems using QGas.

We start from a dataset representing the existing natural gas infrastructure in Carinthia. Planned infrastructure developments are derived from publicly available sources, including the Hydrogen Expansion Roadmap published by the Austrian Gas Grid Management (AGGM)~\cite{AGGM2024H2Roadmap} alongside information on hydrogen production from the HI2-Valley initiative~\cite{HI2Valley2024}. Furthermore, feasibility studies on CO$_2$ transport infrastructure~\cite{AGGM2024CO2Study} are considered to illustrate the integration of additional carriers into the dataset.
The dataset transformation is carried out in several consecutive steps, which are illustrated in Fig.~\ref{fig:Evolution}.

\begin{figure}[t]
    \centering
    \includegraphics[width=1\linewidth]{Images/Case_Studie_Evolution.png}
    \caption{Selected processing steps of the transition in QGas for the generation of the test dataset. Import of relevant infrastructure plans (A), digitization of new pipelines (B), repurposing of existing natural gas pipelines for hydrogen usage (C), and decoupling of the different carrier systems (D).}
    \label{fig:Evolution}
\end{figure}

The process starts by importing hydrogen and CO\textsubscript{2} infrastructure plans in image format into the QGas environment and georeferencing them using identifiable landmarks in the OpenStreetMap background layer. These images serve as visual references and for digitizing planned infrastructure elements.

For hydrogen, most planned infrastructure consists of repurposed natural gas pipelines already present in the dataset with existing attributes such as diameter and pressure. To preserve these attributes, a dedicated hydrogen sublayer is created as a clone of the natural gas pipeline layer, allowing modifications without altering the underlying graph-based network representation. Selected pipeline segments are transferred to the hydrogen sublayer to represent repurposed infrastructure. Minor geometric adjustments are made using the Edit Geometry tool, which allows nodes to be repositioned and pipeline routes modified while preserving the underlying graph topology. Shared nodes between hydrogen and natural gas networks are decoupled using the Split Node tool, which divides nodes while maintaining structural integrity.

For new CO\textsubscript{2} pipelines and production sites, additional infrastructure layers are created using the Add New Element tool before digitalizing them from the semi-transparent infrastructure plans via the Add Infrastructure and Add Pipeline tools. Geometric lengths are calculated automatically, and predefined attribute fields are assigned to ensure dataset consistency across carriers.

\begin{figure}[!t]
    \centering
    \includegraphics[width=1\linewidth]{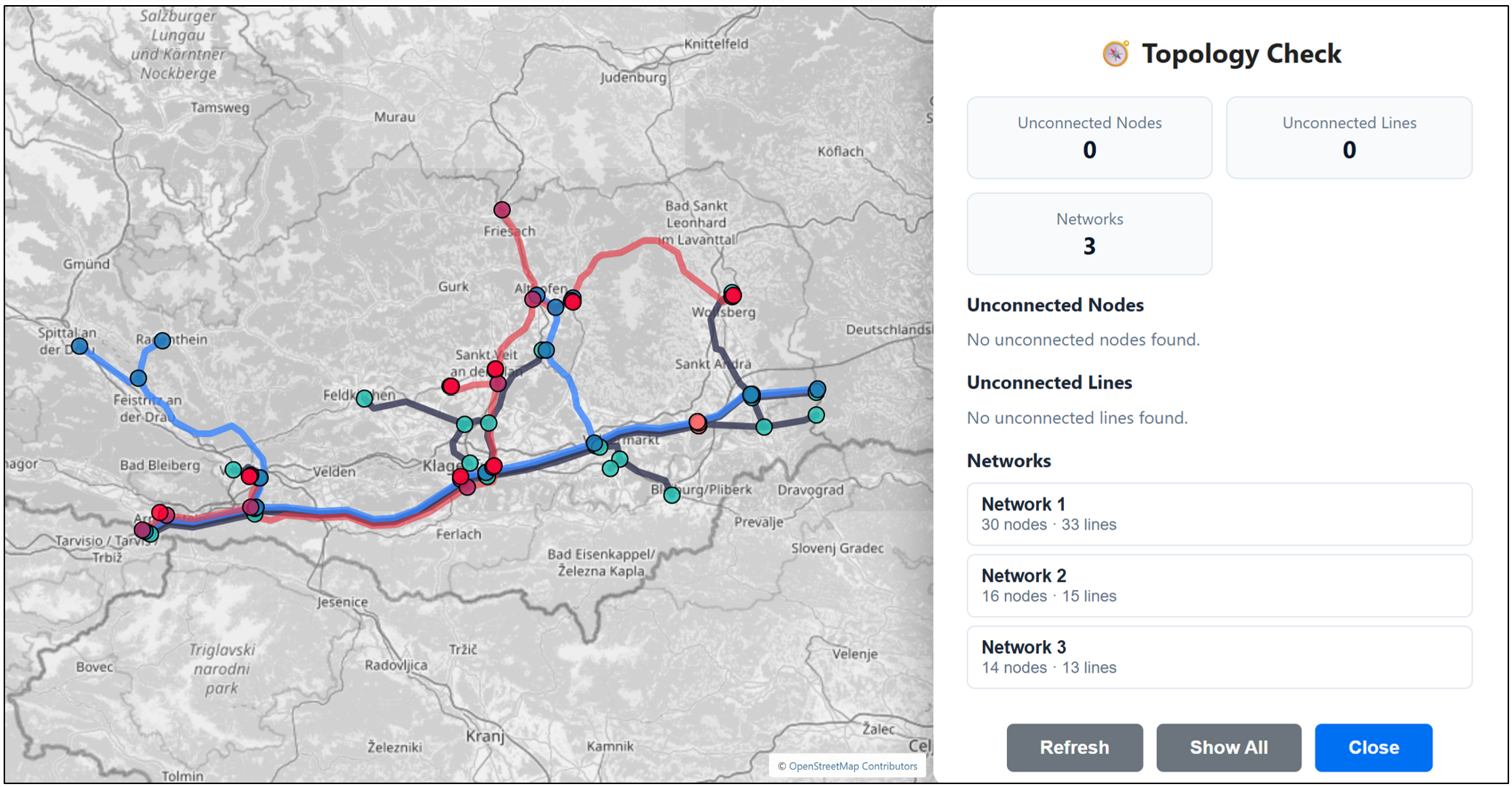}
    \caption{Network verification with the Topology Checker tool.}
    \label{fig:Topology_Check}
\end{figure}

After these structural modifications, the updated network topology is validated using the Topology Checker tool, which detects disconnected nodes, unconnected segments, and independent systems. In this example, the validation process confirms the presence of three distinct networks: natural gas, hydrogen, and CO\textsubscript{2}, as shown in Fig.~\ref{fig:Topology_Check}. Technical attributes such as diameter and pressure levels are then reviewed and updated using the flexible attribute management system in Info Mode, which supports adding, modifying, or removing attributes as needed.

\begin{figure}[!t]
    \centering
    \includegraphics[width=1\linewidth]{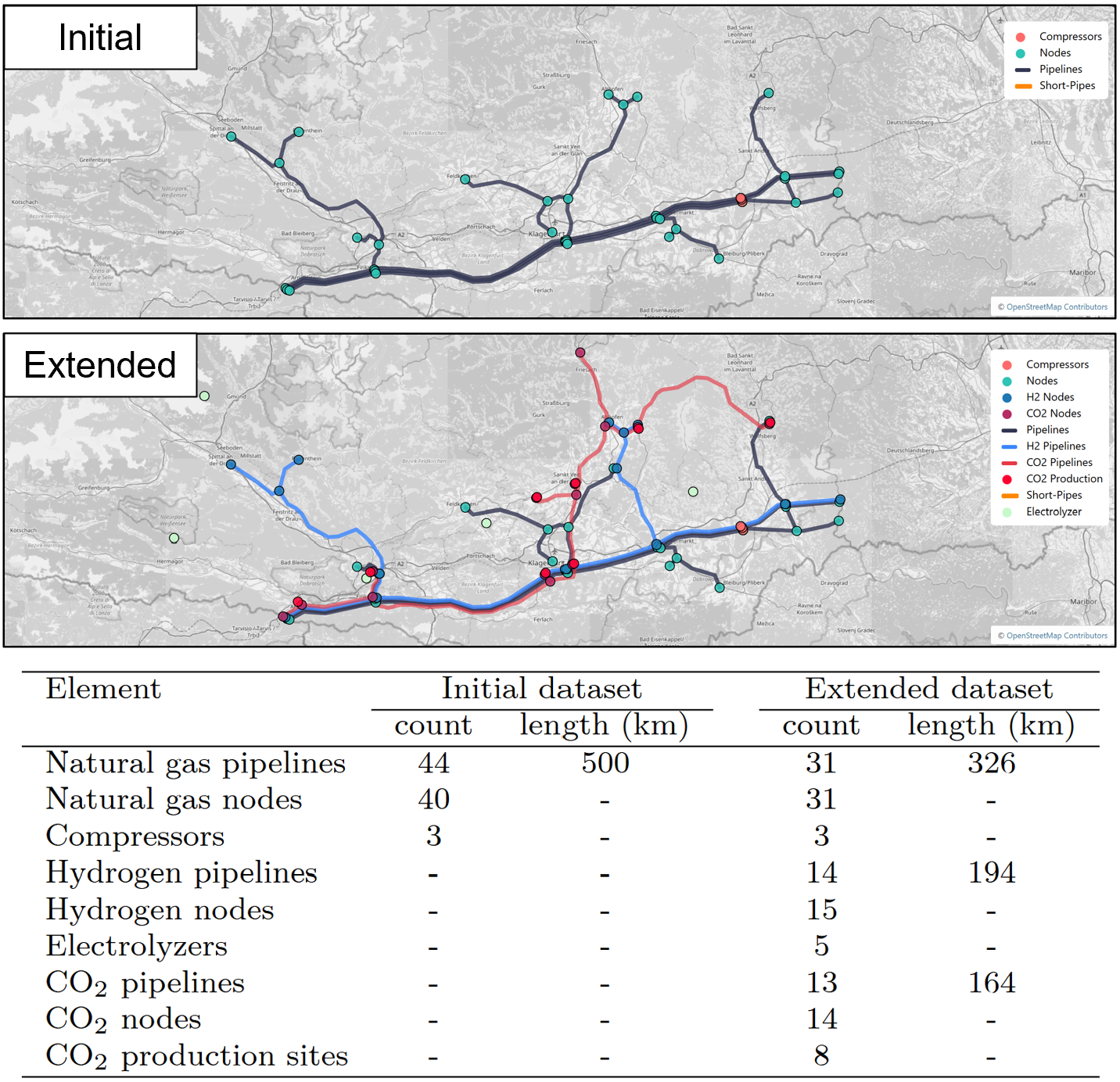}
    \caption{Comparison of the initial and extended dataset.}
    \label{fig:comparison}
\end{figure}

The resulting dataset represents a fully integrated multi-carrier system containing natural gas, hydrogen, and CO\textsubscript{2} networks. A comparison of the initial and modified datasets is shown in Fig.~\ref{fig:comparison}. This example demonstrates how QGas enables transparent, reproducible, and traceable transformation of gas infrastructure datasets, facilitating the preparation of spatially consistent network representations for energy system planning.

\section{Impact}
\label{sec:Impact}
The ongoing decarbonization of the natural gas sector is driving the dynamic evolution of its infrastructure toward a renewable, multi-carrier system integrated with the power sector. To navigate the resulting complexity of modern gas systems, QGas combines the strengths of GIS and graph-based tools, providing users with a practical solution for exploring, expanding, and working with gas infrastructure datasets by enabling consistent, transparent, and reproducible modifications within a unified environment. Crucially, these tasks can be performed through an intuitive graphical user interface, without requiring programming expertise.

In an academic context, QGas has been utilized to generate a visually explorable and easily modifiable version of the well-established SciGrid\_gas dataset~\cite{Pluta2022,Diettrich2021SciGRIDgas}, which is provided as the default dataset for getting familiar with QGas. Beyond existing datasets, QGas facilitates the creation of multi-carrier datasets to support techno-economic studies on hydrogen integration and pipeline repurposing and expansion. To this end, we utilized QGas to compile a holistic dataset~\cite{quantschnig2026dataset} of Austria's natural gas infrastructure and its stage-wise transition towards hydrogen by 2040~\cite{quantschnig2026mapping}. The resulting four-stage dataset provides a coherent basis for use in decision-support tools, such as energy system optimization models. Building on the work of~\cite{quantschnig2025thesis}, we are currently applying QGas to compile a novel, state-of-the-art, European-wide open-source gas infrastructure dataset that aims to close data gaps and overcome inconsistencies in existing counterparts. In the future, we envision QGas as the interface for both gas experts from academia and industry practitioners, enabling them to further enhance the quality of this dataset through their local and sectoral expertise.

Beyond academic research, QGas is an easy-to-use toolkit for stakeholders such as hydrogen project developers, municipalities, and the general public to visually assess gas infrastructure datasets, thereby advancing open access and open data.

\section{Conclusions}
\label{sec:Conclusions}
This paper presented QGas, an interactive toolkit for visualizing, exploring, creating, maintaining, and modifying georeferenced gas infrastructure datasets for energy system planning. By combining GIS-based geometry editing with topology-preserving graph transformations, QGas offers an integrated environment that simplifies dataset development while enhancing consistency, transparency, and reproducibility. An illustrative example demonstrated its applicability for transforming natural gas networks into multi-carrier datasets, including hydrogen and CO$_2$ infrastructures, using a structured workflow for dataset extension. Future developments for QGas will focus on hosting the toolkit as a web-based application, supporting additional data formats, integrating advanced validation and analysis capabilities, and improving interoperability with tools for spatial and network aggregation~\cite{Anarmo2026} and with energy system optimization models.

\section*{Declaration of Generative AI and AI-assisted technologies in the writing process}
During the preparation of this work, the authors used generative AI and AI-assisted technologies. Specifically, GitHub Copilot, supported by large language models including Claude Sonnet 4.5, 4.6, and GPT-5.1 Codex, was employed during the software development process for code structuring, error detection and correction, and code commenting. ChatGPT and Grammarly were used for grammar checking, spell checking, and minor language refinement in both the manuscript and the accompanying technical documentation.

All AI-assisted outputs were carefully reviewed and edited by the authors, who take full responsibility for the content of this publication.

\section*{Acknowledgements}
This work is part of the project iKlimEt (FO999910627), which has received funding in the framework of ”Energieforschung”, a research and technology program of the Klima- und Energiefonds. The authors want to thank D. A. Tejada-Arango for his valuable feedback.

{\footnotesize
{\footnotesize
\bibliographystyle{elsarticle-num}
\bibliography{main}
}
}



\end{document}